\documentclass[conference]{IEEEtran}
\IEEEoverridecommandlockouts
\usepackage{cite}

\usepackage[numbers,sort&compress]{natbib}
\usepackage{lipsum}

\usepackage{amsmath,amssymb,amsfonts}
\usepackage{algorithmic}
\usepackage{graphicx}
\usepackage{textcomp}
\usepackage{xcolor}
\def\BibTeX{{\rm B\kern-.05em{\sc i\kern-.025em b}\kern-.08em
    T\kern-.1667em\lower.7ex\hbox{E}\kern-.125emX}}
\begin{document}

\title{Generative Adversarial Network based Cross-Project Fault Prediction\\

\thanks{}
}

\author{\IEEEauthorblockN{Sourabh Pal}
\IEEEauthorblockA{\textit{dept. of Computer Science and Engineering} \\
\textit{Innopolis University}\\
Innopolis, Russian Federation \\
s.pal@innopolis.university}
}

\maketitle

\begin{abstract}
\newline
Background: The early stage of defect prediction in the software development life cycle can reduce testing effort and ensure the quality of software. Due to the lack of historical data within the same project, Cross-Project Defect Prediction (CPDP) has become a popular research topic among researchers. CPDP trained classifiers based on labeled data sets of one project to predict fault in another project.
\newline
Goals: Software Defect Prediction (SDP) data sets consist of manually designed static features, which are software metrics. In CPDP, source and target project data divergence is the major challenge in achieving high performance. In this paper, we propose a Generative Adversarial Network (GAN)-based data transformation to reduce data divergence between source and target projects.   
\newline
Method: We apply the Generative Adversarial Method where label data sets are choosing as real data, while target data sets are choosing as fake data. The Discriminator tries to measure the perfection of domain adaptation through loss function. Through the generator, target data sets try to adapt the source project domain and, finally, apply machine learning classifier (i.e., Naive Bayes) to classify faulty modules.
\newline
Results: Our result shows that it is possible to predict defects based on the Generative Adversarial Method. Our model performs quite well in a cross-project environment when we choose JDT as a target data sets. However, all chosen data sets are facing a large class imbalance problem which affects the performance of our model.

\end{abstract}


\section{Introduction}
Machine Learning is part of artificial intelligence (AI) study to a build computer algorithms model based on sample data (training data). It is widely used in different domains to make decisions or predictions without giving much effort to programming. These AI algorithms are widely used in different applications such as image processing \cite{b1, b2}, anomaly detection \cite{b3} and so on. The deep learning \cite{b4} approaches are the recent evolution of machine learning. The deep learning applies programmable neural network approaches for decision-making problems, and it can make more accurate decisions with more automation compared to conventional machine learning approaches. It drives the researchers and practitioners to move towards deep learning approaches.

Due to the lack of adequate training data sets, the deep learning model may not perform well. The generative models overcome this problem in two ways: Generative Adversarial Network (GAN) and Variational Autoencoder (VAE). The GAN and VAE differ from each other in terms of architecture. GAN has two components, namely generator and discriminator, whereas VAE has two components, namely encoder and decoder. The generative models can be applied in different areas such as fairness detection using GAN \cite{b5}, Outlier detection using VAE GAN \cite{b6}, and so on. We believe that there are ample opportunities to apply generative models into the SDP domain. In this paper, we will apply GAN-based transfer learning for CPDP model.

SDP approaches aim to guide software developers and testers to identify the most likely defect-prone modules at the early stage of software development with less time and effort compared to manual inspection of source code. With the growing demand for complex systems (e.g., Cyber-Physical Systems, distributed systems, cloud services, etc.), the requirement for software reliability assurance is also growing. Manual maintenance of such complex software systems is difficult, time-consuming, and cost-effective. Therefore, the early stage fault prediction of the software development life cycle may reduce the maintenance cost and increase software quality.

SDP classifies software modules into two categories, namely faulty modules and non-faulty modules based on software metrics \cite{b7}. Software metrics data sets are extracted from historical data of source code repository. The machine learning approaches are the most favored techniques used for the classification in SDP \cite{b8}. Depending on the availability of the data sets, the SDP can be classified into different categories. Within-project Defect Prediction (WPDP) \cite{b9, b10} and Cross-Project Defect Prediction (CPDP) \cite{b11, b12} are the most frequently used categories among them. In WPDP models, the datasets are collected from the same locally available projects where the source and target data sets are homogeneous because both source and target data sets are collected from the same projects. In the early stage of software development projects, it is almost difficult to find enough historical data from the same projects. It makes researchers focus on CPDP. Unlike WPDP, the data sets of CPDP are divergent because source data sets(Training data sets) and target data sets (Test data sets) are extracted from different projects. The Transfer Learning \cite{b13} approaches are the dominating approaches used in CPDP to reduce data divergence between source and target projects. The prediction models based on deep learning are widely used in different areas e.g., image classification \cite{b14}, Sentiment Analysis \cite{b15}, and so on. In this paper, we will focus on Generative Adversarial Network (GAN) based on Transfer Learning to reduce data divergence in CPDP. The main contributions of this paper are the following:

\begin{itemize}
    \item First, we considered source data sets as real data and target data sets as noise data. Next, we used Generator to generate fake source data based on noise target data to reduce data distribution divergence between source and target projects. Finally, we used Naive Bayes classifiers to classify faulty and non-faulty modules where labeled source data sets were used to train the classifier.
    \item Second, we measured how the performance of the model changes with the increasing number of epochs in f-measure.
\end{itemize}

The paper is organized as follows; First, section II introduces the Related work of CPDP. Second, section III describes the Research Methodology. Third, section IV introduces experiments and evaluation measures. Finally, section VI concludes with future work.

\section{Related Work}
The lack of historical data sets at early stages of projects attracts researcher's attention to CPDP \cite{b16, b17, b18, b19}. In this section, we will introduce Statistical and Machine Learning-based CPDP, Deep Learning-based CPDP, and Generative Adversarial Network.

\subsection{Statistical and Machine Learning based CPDP}
Researchers build their prediction models based on software metrics derived from source code repository (e.g., Change metrics \cite{b7}, CK metrics \cite{b20}, Object-oriented metrics \cite{b9}) using machine learning classifiers (e.g., Naive Bayes \cite{b21}, Support Vector Machine \cite{b22}, Decision Tree \cite{b23}, Random Forest \cite{b24}) to classify faulty and non-faulty modules. The main challenge of CPDP is to reduce data divergence between source and target projects data sets. The following studies provided significant contributions to transfer learning in CPDP.

Jaechang Nam et al. \cite{b13} proposed a state-of-the-art transfer learning method, Transfer Component Analysis +(TCA+) to reduce feature distribution between source and target projects. The motivation behind Transfer Component Analysis (TCA) \cite{b25} is to extract common latent factors between source and target projects to bring the source and target data sets into latent space. Let the given source project data set be $D_S = \{(X_{S_i}, y_{S_i})\}_{i=1}^{n_1}$, where $x_{S_i} $ $\subseteq $ $\mathbb{R}^{1 \times m}$ is the input which is represented by a vector of metrics, and $y_{S_i}$ is the corresponding defect information which indicates whether the file is faulty or non-faulty. Lets us assume that target project data sets $D_T = \{X_{T_i}\}_{i=1}^{n_2}$, $X_{S_i}$ and $X_{T_i}$ represent common sets of metrics, and $n_1$ and $n_2$ represent the number of files in source and target projects respectively. 

Let $P(X_S)$ and $P(X_T)$ be the marginal distributions of $X_S = \{X_{S_i}\}_{i=1}^{n_1}$ and $X_T = \{X_{T_i}\}_{i=1}^{n_2}$. The data distribution of $P(X_S)$ and $P(X_T)$ are different in CPDP. The objective of TCA is to learn data transformation for both source and target projects into the latent features space so that the data distribution $\varphi (X_S)$ and $\varphi (X_T)$ becomes small.

Mathematically, TCA tries to map original data of source and target domains onto a latent space where the difference between domains is small $ Dist(\varphi(X_S), \varphi(X_T))$, and data variance of $ Var(\{ \varphi(X_S), \varphi(X_T)\})$ is large. TCA+ is the extension of TCA. In TCA+, the data normalized is based on a set of rules. First, it computes the Euclidean distance for all instances for source and target projects, DIST, as follows:

$DIST = \{d_{ij} : \forall i, j, 1 \leq i < n,1 < j \leq n, i < j\}$

Second, it prepares a set of five rules based on the mean value of Euclidean distance, the median value of Euclidean distance, the minimum value of Euclidean distance, the maximum value of Euclidean distance, and standard deviation of Euclidean distance to choose suitable data normalization to reduce data diversity. The rules are as follows:

\begin{itemize}
    \item \textbf{Rule 1:} If $S_{S\Rightarrow T}[Mean \hspace{0.1cm} value \hspace{0.1cm} of \hspace{0.1cm} DIST]$ and $S_{S\Rightarrow T}[Standard \hspace{0.1cm} deviation \hspace{0.1cm} of \hspace{0.1cm} DIST]$ are individually "SAME", then do not apply any normalization.
    
    \item \textbf{Rule 2:} If $S_{S\Rightarrow T}[Minimum \hspace{0.1cm} value \hspace{0.1cm} of \hspace{0.1cm} DIST]$, $S_{S\Rightarrow T}[Maximum \hspace{0.1cm} value \hspace{0.1cm} of \hspace{0.1cm} DIST]$, and $S_{S\Rightarrow T}[The \hspace{0.1cm} number \hspace{0.1cm} of \hspace{0.1cm} instances]$ are individually either “MUCH LESS” or “MUCH MORE”, then choose min-max normalization.
    
    \item \textbf{Rule 3:} There are two different scenarios. First, $S_{S\Rightarrow T}[Standard \hspace{0.1cm} deviation \hspace{0.1cm} of \hspace{0.1cm} DIST]$ is “MUCH MORE” and the number of instances of a target project is less than that of a source project. Second, $S_{S\Rightarrow T}[Standard \hspace{0.1cm} deviation \hspace{0.1cm} of \hspace{0.1cm} DIST]$ is “MUCH LESS” and the number of instances of a target project is greater than the number of instances in the source project. This situation indicates that target data may have little statistical information. Finally, it is normalized using Mean value of DIST and Standard deviation of DIST.
    
    \item \textbf{Rule 4:} There are two different scenarios. First, $S_{S\Rightarrow T}[Standard \hspace{0.1cm} deviation \hspace{0.1cm} of \hspace{0.1cm} DIST]$ is “MUCH MORE” and the number of instances of a target project is greater than that of a source project. Second, $S_{S\Rightarrow T}[Standard \hspace{0.1cm} deviation \hspace{0.1cm} of \hspace{0.1cm} DIST]$ is “MUCH LESS” and the number of instances of a target project is less than the number of instances in the source project. This situation indicates that target data may have little statistical information. Finally, it is normalized using Mean value of DIST and Standard deviation of DIST.
    
    \item \textbf{Rule 5:} If none of the rules (Rules 1 to 4) are applicable, then choose z-score normalization to make the mean and standard deviation similar between two projects.

\end{itemize}

The proposed approach significantly improves the performance of CPDP, and the performance is also comparable with WPDP.

Chao Liu et al. \cite{b11} proposed a state-of-the-art two-phase transfer learning model (TPTL) to extend the previous work of TCA+. It consists of two phases, namely the project selection phase and the data normalization phase using TCA+. First, they applied source project estimator (SPE) to automatically choose two source projects which had high data distribution similarity with target projects. Second, they generated prediction models based on two source projects by using TCA+ and combining their prediction results to improve the performance of CPDP. By stabilizing TCA+, the model proposed by Chao Liu et al. significantly improved the performance of CPDP.

The SDP is considered a binary classification problem where the classifier classifies the modules into two categories, namely faulty and non-faulty modules. The above-mentioned related works utilized machine learning classifiers to identify faulty modules. In this section, we mentioned the most successful transfer learning approach in CPDP.

\subsection{Deep Learning-based CPDP}
Researchers focus on the application of deep learning in CPDP to explore the semantic features of programs. The traditional machine learning-based models ignore the semantics features of source code. The deep learning-based approaches can learn the semantic features of source code. In this section, we will discuss the contribution of Adversarial Neural Network to reduce data distribution in CPDP.

Lei Sheng et al. \cite{b26} proposed Adversarial Discriminative Convolutional Neural Network (ADCNN) which reduces distribution divergence between source and target projects. It consists of two phases. First, it extracts token vectors from source code Abstract Syntax Trees (ASTs). It chooses three categories and types of AST \cite{b27}, namely nodes of method invocations and class instance creations (e.g., MethodInvocator, SuperMethodInvocator, ClassCreator), declaration nodes (e.g., PackageDeclarator, InterfaceDeclarator and so on), and Control-flow nodes (e.g., ForControl, IfStmt, ForStmt and so on). Second, it uses ADCNN to learn semantic features of both source and target projects using Convolutional Neural Network (CNN), and it maps the target projects onto source projects feature space based on discriminator. Firstly, it trains the parameters of the source encoder and source classifier based on source labeled data. Secondly, it trains the target encoder adversarially so that a discriminator cannot reliably predict the source and target instances to the right domain. The results show that ADCNN model performs well compared to other models (e.g., TCA+).

However, the above mentioned deep learning model did not consider the performance of CPDP model based on only handcrafted features. The above-proposed approach transforms both the source and target projects features into semantic features, and it uses a discriminator to reduce the data distribution gaps between those semantic features. Moreover, it did not consider the model based on transforming only the target project and keeping the source labeled data sets real. In this paper, I applied GAN I kept source data as real and transform target data to reduce data distribution between source and target data.

\subsection{Generative Adversarial Network}
Generative Adversarial Network (GAN) \cite{b28} is a deep learning model which is used to generate fake data. It is also widely used in domain adaptation to minimize data divergence between data sets of different domains. It consists of two networks, namely Generator and Discriminator. The task of the Generator is to generate fake data based on the noise input vector. The task of the discriminator is to learn to represent and identify real data distribution and fake data distribution. The purpose of the training is to minimize the loss of the Generator and maximize the loss of the Discriminator. The minimum loss of the Generator indicates that the generated data is almost real data, and the maximum loss of the discriminator indicates that the discriminator is unable to distinguish between real data and generated data.

In this paper, I applied the Generative Adversarial Network model to reduce data distribution differences between the source and target projects. I used source data as real data and target data as noise vectors and vice versa. Therefore, my model minimized the data distribution difference.

\section{Research Methodology}

This section explores the research questions which are going to analyze in this paper. It also explore the approach applied in this paper. This approaches mainly consist of three steps: Software Metrics Extraction, Build to Train VGAN to generate likely fake data, and constructing Naive Bayes classifier to identify faulty and non-faulty modules.

\subsection{Software Metrics}

The first stage of my model is to extract software metrics from source code. I used Eclipse open source projects data sets for my prediction model \cite{b29}. I used the change metrics (e.g. number Of Versions Until, number Of Fixes Until, number Of Refactoring Until, number Of Authors Until, etc.) \cite{b7}, Chidamber \& Kemerer metrics (e.g. Weighted Methods per Class, Depth of Inheritance Tree of a class, Number Of Children of a Class, Coupling Between Object classes, etc.) \cite{b16}, Object-oriented metrics (e.g. number of methods, number of attributes, number Of Private Attributes, number Of Private Methods, etc.) and so on. Raimund Moser et al. \cite{b7} found that process metrics perform better than product metrics on Eclipse data sets. In this experiment, I used both the product and process metrics for the experiment to analyze the performance of CPDP by using both metrics.

\subsection{VGAN based CPDP Model}

\lipsum[0]
\begin{figure*}
\centering
\includegraphics[height=2.5in, width=6.5in]{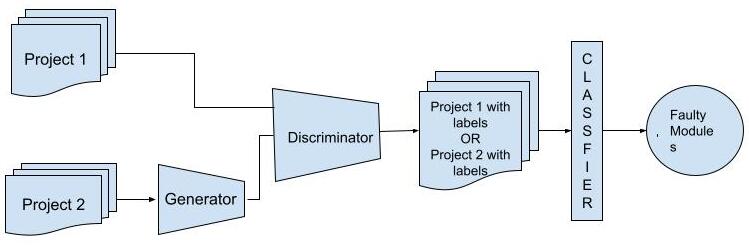}
\caption{VGAN based CPDP Model.}
\end{figure*}

The VGAN \cite{b28} adversarial model is composed of two models: a Generator model G and a Discriminator model D. The Generator is used to capture similar data distribution such as real data, whereas the Discriminator estimates the probability of a sample coming from real data and the generated data. The purpose of the training process of G is to maximize the probability of D. Hence, the training process is to minimize the loss of the Generator and maximize the loss of the Discriminator. However, Generative and Discriminative relationship between source and target projects can be applied in CPDP.

In Figure 1 represent my VGAN-based CPDP model which contains two parts, namely VGAN and Classifier. I experimented with the model by considering project 1 as the source project and project 2 as the target project. I considered the source project (project 1) as real data and the target project (project 2) as noise variables. Hence, the target project was considered as an input vector to the generator to generate a source project such as similar data distribution. The purpose of the Generator is to minimize the loss do that target project data sets becomes similar to the source project. Further, the proposed model trained the classifier to identify faulty modules based on labeled real source data sets. I applied transformed target data sets on the Naive Bayes classifier to identify faulty and non-faulty modules.

I used the fully connected neural network for both Generator and Discriminator and four-layer of neurons for both Generator and Discriminator. 
The output layer of the discriminator employs the sigmoid as the activation function, while the others apply ReLU as the activation function. I obtained 110 features from software engineering data sets and kept all 110 features in source projects. I also used all 110 features of target projects in the generator. Finally, the Generator provided the output with 110 features to make both source and target data feature the similar. In contrast, in Discriminator, I used all 110 features of both source and target data and provided only one output to measure whether the data were real data.  

In the CPDP field, we assume that the source project distribution $p_s(x, y)$ has source data $X_s$ and labels $Y_s$. In contrast, the target project data contain only $p_t(z, k)$ with no labeling. We apply the Naive Bayes classifier to train the classifier using source project data with labels. During the training process, we utilize a minibatch stochastic gradient descent (SGD) algorithm \cite{b30} with Adam optimizer \cite{b31} to optimize the following objective optimization function. We can consider Generator as G and Discriminator as D. The task of G is to make target data similar to source data to minimize the distribution between source and target. In contrast, the task of D is to identify whether data are fake by maximizing the objective function. So, D and G can be written as min-max function as follows(equation 1):

\begin{equation}
    \begin{split}
        \min_G\max_D V(D, G) &= \mathbb{E}_{x\sim p_s(x)}[log D(x)] + \\
        & \mathbb{E}_{z\sim p_t(z)}[log(1 - D(G(z)))]
    \end{split}
\end{equation}

\section{Experiments}
This section discusses the data sets and the evaluation measures. It also outlines the four Eclipse projects which I used for the experiment. To be more precise, I used Eclipse open source projects data sets for my prediction model \cite{b29}. I choosed three projects from Eclipse, namely equinox, JDT, and lucene. Table 1 provides detailed information on the selected projects. I choose equinox as the source project to train the model, while JDT and lucene were used as the target project to predict the performance of model. Finally, We measured the performance of the model in terms of f1-measure.

\begin{table}
\begin{center}
\caption{Project Details}
\begin{tabular}{c c c c} 
\hline
\hline
Project Name & Number of Instances & Buggy Rate\\ 
\hline\hline
equinox & 324 & 39.81 \\
\hline
JDT & 997 & 20.66 \\
\hline
lucene & 691 & 9.26 \\
\hline
\end{tabular}
\end{center}
\end{table}

\section{Results}

This section answer two research questions proposed in the previous section. I choose three open-source Eclipse projects for experimentation. Eclipse projects contain many sub-projects. I choose three projects among them namely equinox, JDT, and lucene for the experiment. I choose equinox as the source project, whereas I used JDT and lucene as the target project. 
By doing this, I was able to generate two different conditions for the experiment: eclipse as source project with JDT as target project and eclipse as source project and lucene as target project. Then I calculated F-measure to measure the performance of both conditions. I also calculated the F-measure performance for four different epochs i.e., 25, 50, 75, and 100. In Figure 2 shows the performance of both experimental conditions. In general, it shows that equinox as a source project works well to predict defects in the JDT project compare to lucene project. It also demonstrate the performance of the prediction model does not change much with the increasing number of epochs.

\begin{figure}
\centering
\includegraphics[height=2.5in, width=3.5in]{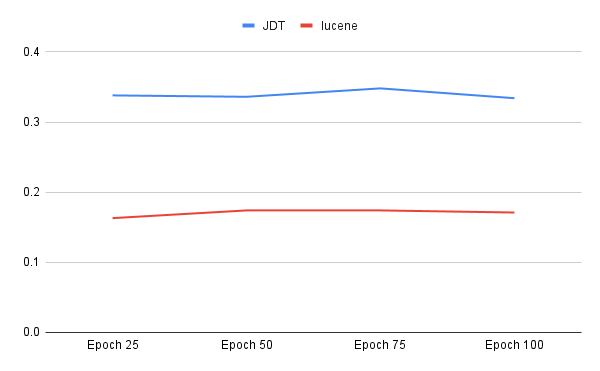}
\caption{Performance of VGAN based CPDP Model.}
\end{figure}

\section{Conclusion and Future Work}

In this paper, we tried to propose an approach to apply the domain adaptation approach for CPDP. My VGAN-based CPDP approach utilizes the fully connected neural network for both generator and discriminator for domain adaptation. However, JDT as target projects gives reasonably good result compared to lucene used as target projects. I measured the performance of the F-measure because I wanted to explore whether my model is perfect enough to be applied in real time. It is clear from Table 1 that my data sets have class imbalance problem. In lucene data sets, only 9.26 percent of data sets contain buggy. In the future, I intent to solve the class imbalance problem in the data sets and do experiments with larger data sets to improve the performance of the proposed model.

\end{document}